\documentclass[12pt,a4paper]{article}
\usepackage{mathrsfs}
\usepackage{epsfig}
\begin{document}
\title{ Vector bileptons and the decays $h\rightarrow \gamma\gamma,Z\gamma$}
\author{Chong-Xing Yue,  Qiu-Yang Shi, Tian Hua \\
{\small Department of Physics, Liaoning  Normal University, Dalian
116029, P. R. China}
\thanks{E-mail:cxyue@lnnu.edu.cn}}
\date{\today}

\maketitle
\begin{abstract}
Takeing into account of the constraints on the relevant parameters from the muon anomalous magnetic moment,  we  consider the contributions of the vector bileptons $V^{\pm}$ and $U^{\pm\pm}$ predicted by the reduced minimal 331 model to the Higgs decay channels $h \rightarrow \gamma \gamma$ and $Z\gamma$. Our numerical results show that  the vector bileptons can enhance the partial width  $\Gamma(h\rightarrow \gamma\gamma)$, while reduce the partial width  $\Gamma(h\rightarrow Z\gamma)$, which are  anti-correlated. With reasonable values of the relevant free parameters, the vector bileptons can explain the $LHC$ data for the $\gamma\gamma$ signal. If the  $CMS$ data persists, the values of the free parameters $ \lambda_{2}$ and $ \lambda_{3}$ should be severe constrained.

\vspace{0.5cm} \vspace{2.0cm} \noindent

\end{abstract}
\newpage
\noindent{\bf 1. Introduction }\vspace{0.5cm}

The exciting $LHC$ discovery of a Higgs-like boson of mass around 125$GeV$ [1] seems to experimentally complete the standard model ($SM$) of particle physics. In spite of successes of the $SM$, there are profound experimental and theoretical reasons to suppose that it is a low-energy effective theory. The small but non-vanishing neutrino masses, the hierarchy and naturalness problems provide a strong motivation for contemplating new physics beyond the $SM$ at $TeV$ scale.

 All new physics models have in common the prediction of new particles. Bileptons are one kind of new particles, which carry two units of lepton number and which couple to two leptons, but not to the $SM$ quarks. They are present in several new physics scenarios, such as left-right symmetric models, composite and technicolor theories. Vector bileptons in which we are interested are massive gauge bosons, which appear in extended gauge models, where the electroweak gauge group is imbedded in a larger group, as the $SU(5)$ grand unified theory [2] and the so-called 331 models [3]. Production of bileptons in high-energy collider experiments has been extensively studied in Refs.[4, 5, 6].

Although  bileptons have not been found yet experimentally, but it is widely believed that the $LHC$ is able to probe them in the coming years. The $LEP$ II experiments have  given the lower limits on the vector bilepton masses around $100 GeV$ [7]. The stringent  lower bounds on vector bileptons require these particles to be heavier than $850GeV$ and $ 750 GeV$, which were established from muonium-antimuonium conversion [8] and from fermion pair production and lepton-flavor violating processes [9], respectively. Both $ATLAS$ and $CMS$ collaborations have searched for long-lived doubly-charged states [10], the lower limit about $680GeV$ is obtained from the $LHC$ data. All these existing mass bounds can however be easily evaded by making less restrictive assumptions than the aforementioned analyses.

The stringent lower bounds on vector bileptons are obtained from the assumption that the bilepton coupling is flavor-diagonal. It is well known that the 331 model can generally induce flavor mixing and its extended Higgs sector has contributions to muonium to antimuonium conversion. Thus the vector bilepton masses can be as low as $350GeV$, as argued by Ref.[11]. Recently, Ref.[12] assumed  that the couplings of the doubly-charged gauge bosons to the $SM$ leptons are flavor-conserving and model  independent discuss  their production and signatures at the $LHC$. It is found that their masses can be reduced  to $400GeV$, which might still be  satisfy the $LHC$ constraints. We hence assume the vector-bilepton mass as a free parameter in this paper.

Considering the constraints of the deviation from the $SM$ prediction of the muon anomalous magnetic moment (muon $g-2$) on the free parameters, we will calculate the contributions of vector bileptons to the Higgs decay channels $h\rightarrow \gamma \gamma$ and $Z \gamma$ in this paper. In general, the couplings of vector bileptons to Higgs boson $h$, photon $\gamma$ and the gauge boson $Z$ are model dependent. To simply our calculation, we restrict ourselves to the reduced minimal 331 ($RM$331) model [13], which is built with only two scalar triplets and can easily give all the scalar-gauge boson and scalar-fermion couplings without making the usual cumbersome assumptions for the couplings in the scalar potential of the original 331 models [14]. We find that the deviation of the measured value for the muon $g-2$ from its $SM$ prediction can indeed give constraints on the masses of the vector bileptons. These new particles can produce significant contributions to the partial widths  $\Gamma(h\rightarrow \gamma \gamma)$ and $\Gamma(h\rightarrow $Z$ \gamma)$, which might explain the $LHC$ data about the $SM$-like Higgs boson.

This paper is organized as follows. In section 2, we review the basics content of the $RM$331 model and give the relevant couplings of vector bileptons to ordinary particles, which are related our calculation. The contributions of vector bileptons to the muon $g-2$ $a_{\mu}$ and the Higgs decay channels $h\rightarrow \gamma \gamma$ and $Z\gamma$ are calculated in sections 3 and 4. Our conclusion and simply discussion are given in section 5.

\vspace{0.5cm} \noindent{\bf 2. The  basics content of the $RM$331 model }

\vspace{0.5cm}The 311 models [3] are based on the gauge symmetry $SU(3)_{C}\times SU(3)_{L}\times U(1)_{X}$, in which the electric charge operator is defined as
\begin{eqnarray}
{Q}=T_{3}+\beta T_{8}+XI,
\end{eqnarray}
where $T_{3}$ and $T_{8}$ are two of eight generators of $SU(3)_{L}$, $X$ is the new quantum number corresponding to $U(1)_{X}$, the free parameter $\beta$ defines the different representation contents and is used to label the particular type of the 331 models and $\beta=-\sqrt{3}$ corresponds to the minimal 331 model [14].

The fermion sector of the $RM$331 model [13] is same as that of the minimal 331 model. The left-handed  fermions transform under the $SU(3)_{L}$ gauge group as the triplets
\begin{eqnarray}
{f_{aL}}=\left(
          \begin{array}{c}
            \nu_{la} \\
            l_{a} \\
            l^{c}_{a} \\
          \end{array}
        \right)_{L}
          \sim
          \left(
                \begin{array}{ccc}
                  1 ,& 3,& 0 \\
                \end{array}
              \right),
\end{eqnarray}
\begin{eqnarray}
{Q_{1L}}=\left(
          \begin{array}{c}
            u_{1} \\
            d_{1} \\
            T \\
          \end{array}
        \right)_{L}
          \sim
          \left(
            \begin{array}{ccc}
              3, & 3, & \frac{2}{3} \\
            \end{array}
          \right),
{Q_{iL}}=\left(
          \begin{array}{c}
            d_{i} \\
            -u_{i} \\
            D_{i} \\
          \end{array}
        \right)_{L}
          \sim
          \left(
            \begin{array}{ccc}
              3, &3^{\ast},  & -\frac{1}{3} \\
            \end{array}
          \right).
\end{eqnarray}
Where $a=1,2,3$, $i=2,3$, and the number in parenthesis represent the field's transformation properties under the gauge group $SU(3)_{C}\times SU(3)_{L}\times U(1)_{X}$. The new quark $T$ carries electric charge $5/3$ in unit of positron's electric charge, while $D_{2}$ and $D_{3}$ carry electric charge $-4/3$ each one. The right-handed quarks are singlets of the $SU(3)_{L}$ group,
\begin{eqnarray}
{U_{iR}^{a}}\sim\left(
                 \begin{array}{ccc}
                   3 ,& 1 ,& \frac{2}{3} \\
                 \end{array}
               \right),&&
{d_{iR}^{a}}\sim\left(
                 \begin{array}{ccc}
                   3 ,& 1 ,& -\frac{1}{3} \\
                 \end{array}
               \right),
\end{eqnarray}
\begin{eqnarray}
{T_{R}}\sim\left(
                 \begin{array}{ccc}
                   3 ,& 1 ,& \frac{5}{3} \\
                 \end{array}
               \right),&&
{D_{iR}}\sim\left(
                 \begin{array}{ccc}
                   3 ,& 1 ,& -\frac{4}{3} \\
                 \end{array}
               \right).
\end{eqnarray}

For the $RM$331 model, the scalar sector contains only two Higgs triplets, which is different from that of the minimal 331 model,
\begin{eqnarray}
{\rho}=\left(
         \begin{array}{c}
           \rho^{+} \\
         \rho^{0}\\
           \rho^{++}\\
         \end{array}
       \right)\sim\left(
                    \begin{array}{ccc}
                      1, & 3, & 1 \\
                    \end{array}
                  \right),&&
{\chi}=\left(
         \begin{array}{c}
           \chi^{-} \\
        \chi^{--}\\
          \chi^{0}\\
         \end{array}
       \right)\sim\left(
                    \begin{array}{ccc}
                      1, & 3, & -1 \\
                    \end{array}
                  \right).
\end{eqnarray}
The most general gauge and Lorentz invariant scalar potential is given by [13]
\begin{eqnarray}
{V(\mu,\rho)}=\mu^{2}_{1}\rho^{+}\rho+\mu^{2}_{2}\chi^{2}\chi+ \lambda_{1}(\rho^{+}\rho)^{2}+ \lambda_{2}(\chi^{+}\chi)^{2}\nonumber\\
+\lambda_{3}(\rho^{+}\rho)(\chi^{+}\chi)+ \lambda_{4}(\rho^{+}\chi)(\chi^{+}\rho).
\end{eqnarray}
The triplet $\chi$ governs the symmetry breaking of $SU(3)_{L} \times U(1)_{X}$ down to $SU(2)_{L} \times U(1)_{Y}$, while $\rho$ is responsible for breaking $SU(2)_{L} \times U(1)_{Y} \rightarrow U(1)_{em}$. The neutral scalars develop the vacuum expectation values (VEVs) $\langle \rho^{0} \rangle=\nu_{\rho} /\sqrt{2}$ and $\langle \chi^{0} \rangle=\nu_{\chi}/\sqrt{2}$ with $\nu_{\chi}\gg\nu_{\rho}$ and $\nu_{\rho}\approx\nu=246GeV$. After the symmetry breaking, the masses of the physical scalars $H^{\pm\pm}$, $H_{1}$ and $H_{2}$ are

\begin{eqnarray}
{M_{H^{\pm \pm}}^{2}}=\frac{\lambda_{4}}{2}(\nu_{\chi}^{2}+\nu_{\rho}^{2}),&
{M_{H_{1}}^{2}}=(\lambda_{1} - \frac{\lambda_{3}^{2}}{4\lambda_{2}})\nu_{\rho}^{2},&
{M_{H_{2}}^{2}}=\lambda_{2}\nu_{\chi}^{2} + \frac{\lambda_{3}^{2}}{4 \lambda_{2}} \nu_{\rho}^{2}.
\end{eqnarray}
Where the lightest neutral scalar $H_{1}$ is a $SU(2)_{L}$ component and can be identified as the $SM$-like Higgs boson $h$, the dimensionless free parameters $\lambda_{i}(i=1,2,3,$ and $4)$ satisfy the relations: $\lambda_{1}>0, \lambda_{2}>0, \lambda_{3}<0$ and $4\lambda_{1}\lambda_{2}>\lambda_{3}^{2}$  [13].

Similar with the minimal 331 model, the $RM$331 model predicts the single-charged ($V^{\pm}$) and doubly-charged ($U^{\pm\pm}$) vector bileptons and neutral boson $Z^{'}$ in addition to the electroweak gauge bosons $W^{\pm}$ and $Z$. The interactions of the Higgs bosons with these gauge bosons are described by
\begin{eqnarray}
{\cal
L}=(D_{\mu}\chi)^{+}(D^{\mu}\chi)+(D_{\mu}\rho)^{+}(D^{\mu}\rho),
\end{eqnarray}

where
\begin{eqnarray}
{D_{\mu}}=\partial_{\mu}-igA_{\mu}^{a}\frac{\lambda^{a}}{2}-ig_{X}X\frac{\lambda_{9}}{2}B_{\mu}
\end{eqnarray}
with $\lambda^{a}(a=1,\cdot\cdot\cdot,8)$ being the Gellmann matrices and $\lambda_{9}=\sqrt{\frac{2}{3}}$  diag  $\left(
                                                                                                                   \begin{array}{ccc}
                                                                                                                     1, & 1, & 1 \\
                                                                                                                   \end{array}
                                                                                                                 \right)
$. From above equations one can obtain the eigenstates of the charged gauge bosons and their respective masses:
\begin{eqnarray}
W^{\pm}=\frac{A^{1}\mp iA^{2}}{2} \longrightarrow  m_{W}^{2}=\frac{g^{2} \nu_{\rho}^{2}}{4},
\end{eqnarray}
\begin{eqnarray}
V^{\pm}=\frac{A^{4}\pm iA^{5}}{\sqrt{2}}\longrightarrow M_{V}^{2}=\frac{g^{2}\nu_{\chi}^{2}}{4},
\end{eqnarray}
\begin{eqnarray}
U^{\pm\pm}=\frac{A^{6}\pm iA^{7}}{\sqrt{2}}\longrightarrow M_{U}^{2}=\frac{g^{2}(\nu_{\rho}^{2}+\nu_{\chi}^{2})}{4}.
\end{eqnarray}
It is obvious that there is $M_{U}^{2}-M_{V}^{2}=m_{W}^{2}$.

The couplings of the vector bileptons $V^{\pm}$ and $U^{\pm\pm}$ to ordinary particles, which are related to our calculation, can be written as
\begin{eqnarray}
g^{V\nu_{\mu}\mu}=\frac{e}{\sqrt{2}S_{W}}V_{PMNS}, && g^{U\mu\mu}=\frac{e}{\sqrt{2}S_{W}};
\end{eqnarray}
\begin{eqnarray}
g^{H_{1}VV}=-\frac{e^{2}}{2S_{W}^{2}}S_{\beta}\nu_{\chi},&& g^{H_{1}UU}=\frac{e^{2}}{2S_{W}^{2}}(C_{\beta}\nu_{\rho}-S_{\beta}\nu_{\chi});
\end{eqnarray}
\begin{eqnarray}
g^{\gamma VV}=e,&& g^{\gamma UU}=2e;
\end{eqnarray}
\begin{eqnarray}
g^{ZVV}=-\frac{e(1+2S_{W}^{2})}{2S_{W}C_{W}}, && g^{ZUU}=\frac{e(1-4S_{W}^{2})}{2S_{W}C_{W}}.
\end{eqnarray}
Where $S_{W}=\sin \theta_{W}$, $C_{W}=\cos \theta_{W}$, $\theta_{W}$ is the Weinberg angle, and $V_{PMNS}$ is the mixing matrix. In the case of $\nu_{\chi}\gg\nu_{\rho}$, the mixing parameters $S_{\beta}$ and $C_{\beta}$ satisfy
\begin{eqnarray}
C_{\beta}\approx 1-\frac{\lambda_{3}^{2}}{8\lambda_{2}^{2}}t^{2},S_{\beta}=\frac{\lambda_{3}t}{2\lambda_{2}}
\end{eqnarray}
with $t=\nu_{\rho}/\nu_{\chi}$.

In the following two sections we will use above equations to discuss the contributions of $V^{\pm}$ and $U^{\pm\pm}$ to the muon $g-2$ $a_{\mu}$ and the Higgs decay channels $h\rightarrow \gamma \gamma$ and $Z\gamma$.

\vspace{0.5cm} \noindent{\bf 3. Vector bileptons and the muon $g-2$  $a_{\mu}$ \hspace*{0.6cm} }

\vspace{0.5cm} \hspace{0.1cm}

\begin{figure}[htb]
\vspace{0.5cm}
\begin{center}
 \epsfig{file=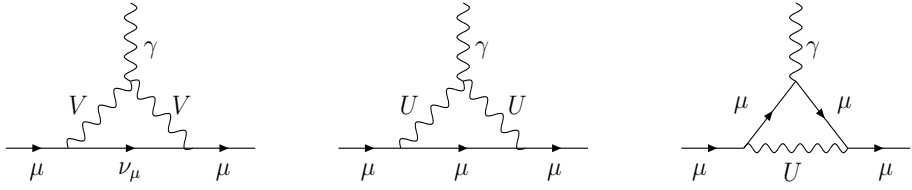,width=340pt,height=70pt}
\vspace{0.2cm} \caption{Leading order diagrams for the contributions of the vector bileptons $V^{\pm}$ \hspace*{1.9cm} and $U^{\pm\pm}$ to the muon $g-2$ $a_{\mu}$.} \label{ee}
\end{center}
\end{figure}
The muon $g-2$  $a_{\mu}$ is one of the most precise measured observables in particle physics [15]. The theoretical prediction from the $SM$ has suggested that there is a discrepancy  between the experimental result and the $SM$ prediction at about $3\sigma$ level [16]
\begin{eqnarray}
\delta a_{\mu}=a_{\mu}^{\exp}-a_{\mu}^{SM}=(26.1\pm8.0)\times10^{-10}.
\end{eqnarray}
If the discrepancy can not be explained in the $SM$, it can be regarded as an evidence of the new physics beyond the $SM$, which should yield the net positive contribution. It is worth considering the new physics seriously in order to explain the anomaly or give constraints on the specific new physics model.

From section 2 we can see that the single-charged and doubly-charged bileptons $V^{\pm}$ and $U^{\pm\pm}$ predicted by the $RM$331 model can contribute to the muon $g-2$ via the Feynman diagrams shown in Fig.1. Since there is $m_{\mu}\ll M_{V}$ or $M_{U}$, in Fig.1 we have dropped any sub-leading diagrams, such as vacuum polarization diagrams. The contributions of the vector bileptons $V^{\pm}$ to $a_{\mu}$ can be written as
\begin{eqnarray}
\delta a_{\mu}^{V}=\frac{g^{2}}{64\pi^{2}} \frac{m_{\mu}^{2}}{M_{V}^{2}} |V_{PMNS}^{\mu\mu}|^{2}F(a),
\end{eqnarray}
where $a=m_{\nu_{\mu}}/M_{V}$ and $F(a)$ is given by
\begin{eqnarray}
F(a)\approx\frac{10-43a^{2}+78a^{4}}{3(1-a^{2})^{4}}\approx\frac{10}{3}.
\end{eqnarray}

The calculation of Fig.1(b) leads to
\begin{eqnarray}
\delta a_{\mu 1 }^{U}=\frac{5g^{2}}{12\pi^{2}} \frac{m_{\mu}^{2}}{M_{U}^{2}}.
\end{eqnarray}
Finally, the contributions of the vector bileptons $U^{\pm\pm}$ to $a_{\mu}$ via Fig.1(c) are given by
\begin{eqnarray}
\delta a_{\mu2}^{U}=\frac{g^{2}}{4 \pi^{2}} \frac{m_{\mu}^{2}}{M_{U}^{2}} F_{2}(x),
\end{eqnarray}
\begin{eqnarray}
F_{2}(x)=\int_{0}^{1}dx \frac{2x-2x^{2}+b^{2}x^{3}}{1-x+b^{2}x^{2}}
\end{eqnarray}
with $b=m_{\mu}/M_{U}$. If we take $b\approx0$, then there is $F_{2}(x)\approx1$. Summating Eq.(20), Eq.(22), and Eq.(23), we can obtain the total contributions of the vector bileptons $V^{\pm}$ and $U^{\pm\pm}$ to the muon $g-2$ $a_{\mu}$, which agree with the results given by Ref.[17].

\begin{figure}[htb]
\vspace{-0.5cm}
\begin{center}
 \epsfig{file=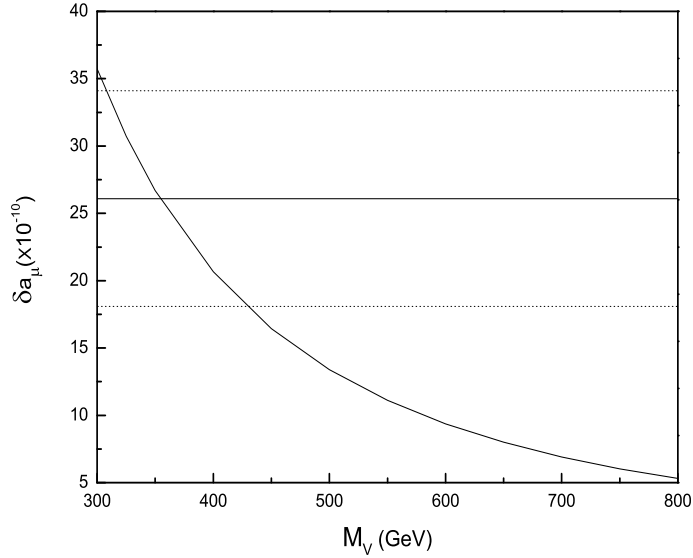,width=300pt,height=260pt}
 \vspace{-0.5cm}
 \caption{ The contributions of the vector bileptons $V^{\pm}$ and $U^{\pm\pm}$ to the muon $g-2$ $a_{\mu}$. The \hspace*{1.7cm} region between dashed lines  corresponds $1\sigma$\ allowed region from $a_{\mu}$ constraints.} \label{ee}
\end{center}
\end{figure}

If the deviation of the measured value for the muon $g-2$ $a_{\mu}$ from its $SM$ prediction indeed exist and is interpreted as signal of new physics, then the discrepancy $\delta a_{\mu}$ should give constraints on the mass parameters related to the vector bileptons $V^{\pm}$ and $U^{\pm\pm}$. The contributions of $V^{\pm}$ and $U^{\pm\pm}$ to $a_{\mu}$ are shown in Fig.2, which shows that the deviation $\delta a_{\mu}$ demands $308GeV \leq M_{V} \leq 429GeV$ at one $\sigma$ level. The muon $g-2$ can give more strict constraint on the lower limit for the mass $M_{V}$ than those given by Ref.[13]. Certainly, if one demands that vector bileptons explain the deviation $\delta a_{\mu}$ at $2\sigma$ level, the limit can be relaxed. Considering this constraint we will calculate the contributions of the vector bileptons $V^{\pm}$ and $U^{\pm\pm}$ to the Higgs decay channels $h \rightarrow \gamma \gamma$ and $Z\gamma$ in the following section.

\vspace{0.5cm} \noindent{\bf 4. Vector bileptons and the $h\rightarrow \gamma\gamma$ and $Z\gamma$ decays \hspace*{0.6cm}  }

\vspace{0.5cm}

The $ATLAS$ and $CMS$ collaborations at the $LHC$ have discovered a Higgs-like particle with mass around $125GeV$ [1]. Both collaborations have reported that the observed diphoton signal strength is sizably larger than the $SM$ prediction. The latest results of the $ATLAS$ and $CMS$ have been announced in Morioned conference [18] and confirm the Higgs discovery with mass of order $125GeV$. The $ATLAS$ still reports significant excesses, $\sigma/\sigma_{M}=1.65\pm0.35$ in the $\gamma\gamma$ channel, while the $CMS$ changes their previous results to $\sigma/\sigma_{M}=1.11\pm0.31$ for the cut-based analysis. It is obvious that further analysis is needed to reveal the discrepancy between the results of the two collaborations. If the enhancement in the diphoton channel persists, it would provide a good indication of new physics.

It is well known that the decay $h\rightarrow \gamma \gamma$ is induced at the loop level, in the $SM$ the $W$-loop contribution is dominant and the top-loop effect is destructive against the $W$-loop [19], and is very sensitive to the new charged partides [20]. These new particles might explain the "excess" in the $\gamma\gamma$ signal strength of the Higgs discovered at the $LHC$ [21, 22]. Due to the electroweak gauge symmetry, any new charged particles affecting the decay $h\rightarrow \gamma\gamma$ can also generate contributions to the decay $h\rightarrow Z\gamma$, which is also a loop-induced process. A combined analysis of $h\rightarrow\gamma\gamma$ and $Z\gamma$ could provide more valuable information about the structure of new physics [23]. On the experimental side, the $CMS$ collaboration has given their crude results of the measurement of $h\rightarrow Z\gamma$ and set an upper limit on $\sigma_{Z\gamma}/\sigma_{Z\gamma}^{SM}<10$ [24].

According the general formula given by Ref.[20], the partial widths of $h \rightarrow \gamma \gamma$ and $h \rightarrow Z\gamma$ including the vector bilepton contributions can be written as
\begin{eqnarray}
\Gamma(h\rightarrow\gamma\gamma)
=\frac{\alpha^{2} m_{h}^{3}}{256\pi^{3}\nu_{\rho}^{2}}[A_{1}(\tau_{W})+\frac{4}{3}
A_{\frac{1}{2}}(\tau_{t}) \nonumber \\-\frac{S_{\beta}m_{W}^{2}}{M_{V}^{2}}\frac{\nu_{\chi}}{\nu_{\rho}}A_{1}(\tau_{V})
+\frac{4m_{W}^{2}}{M_{U}^{2}}(C_{\beta}-S_{\beta}\frac{\nu_{\chi}}{\nu_{\rho}})A_{1}(\tau_{U})]^{2},
\end{eqnarray}
\begin{eqnarray}
\Gamma(h\rightarrow Z\gamma)=
\frac{G_{F}\alpha^{2}m_{h}^{3}}{64\sqrt{2}\pi^{3}}(1-\frac{m_{Z}^{2}}{m_{h}^{2}})^{3}\{\frac{1}{S_{W}}[C_{W}A_{1}(\tau_{W},\lambda_{W})
+\frac{2(1-\frac{8}{3}S_{W}^{2})}{C_{W}}A_{\frac{1}{2}}(\tau_{t},\lambda_{t})]\nonumber\\ +\frac{S_{\beta}m_{W}^{2}}{M_{V}^{2}}\frac{\nu_{\chi}}{\nu_{\rho}}\frac{1+2S_{W}^{2}}{2S_{W}C_{W}}A_{1}(\tau_{V},\lambda_{V})
+\frac{2m_{W}^{2}}{M_{U}^{2}}(C_{\beta}-S_{\beta}\frac{\nu_{\chi}}{\nu_{\rho}})\frac{1-4S_{W}^{2}}{2S_{W}C_{W}}A_{1}(\tau_{U},\lambda_{U})\}^{2},
\end{eqnarray}
where $\tau_{i}=4m_{i}^{2}/m_{h}^{2}$ and $\lambda_{i}=4m_{i}^{2}/m_{Z}^{2}$.
$A_{1}(\tau_{i}), A_{\frac{1}{2}}(\tau_{t}),A_{\frac{1}{2}}(\tau_{t},\lambda_{t})$ and $A_{1}(\tau_{i},\lambda_{i})$ are loop functions defined in Ref.[25].
\begin{figure}[htb]
\vspace{-0.5cm}

{\includegraphics[scale=0.68]{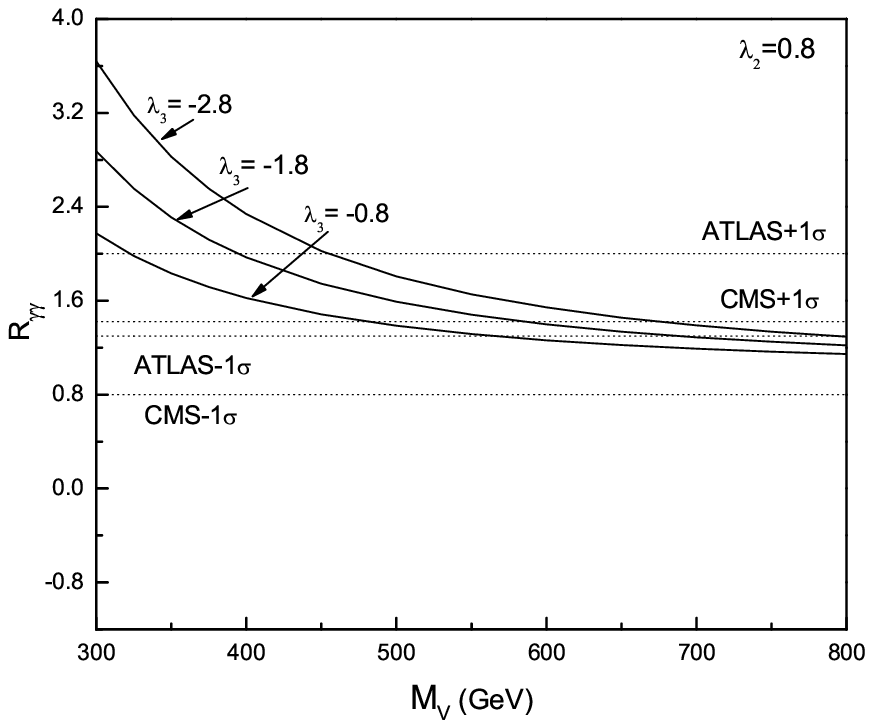}}
{\includegraphics[scale=0.68]{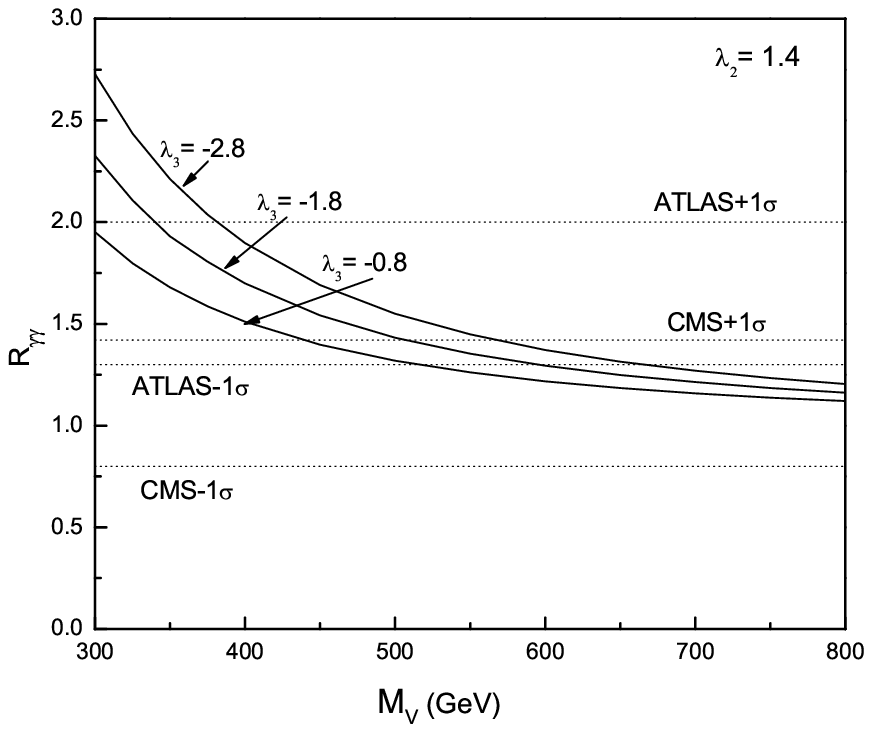}}

{\includegraphics[scale=0.68]{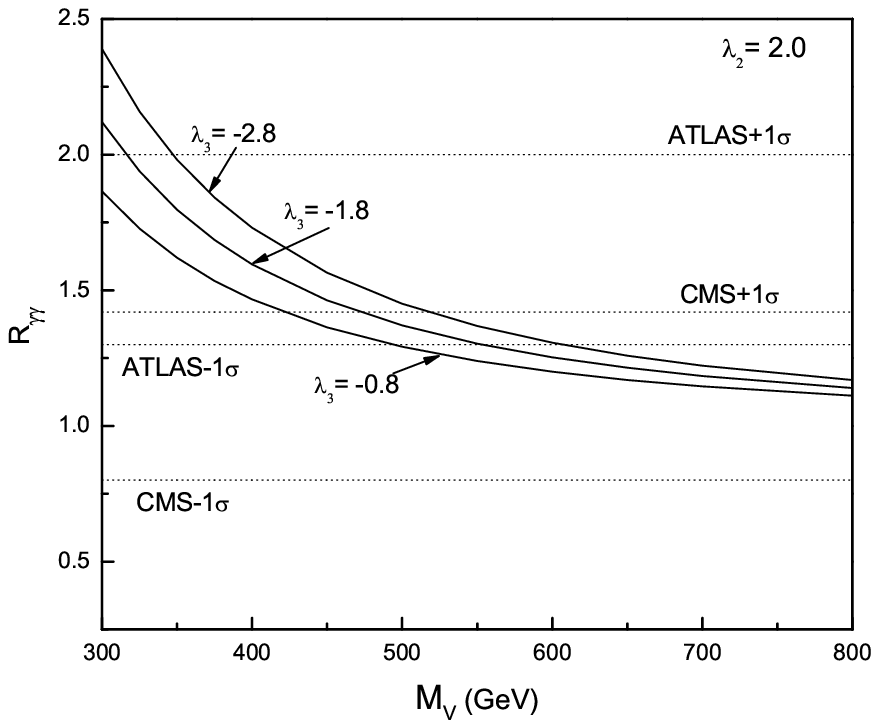}}
{\includegraphics[scale=0.68]{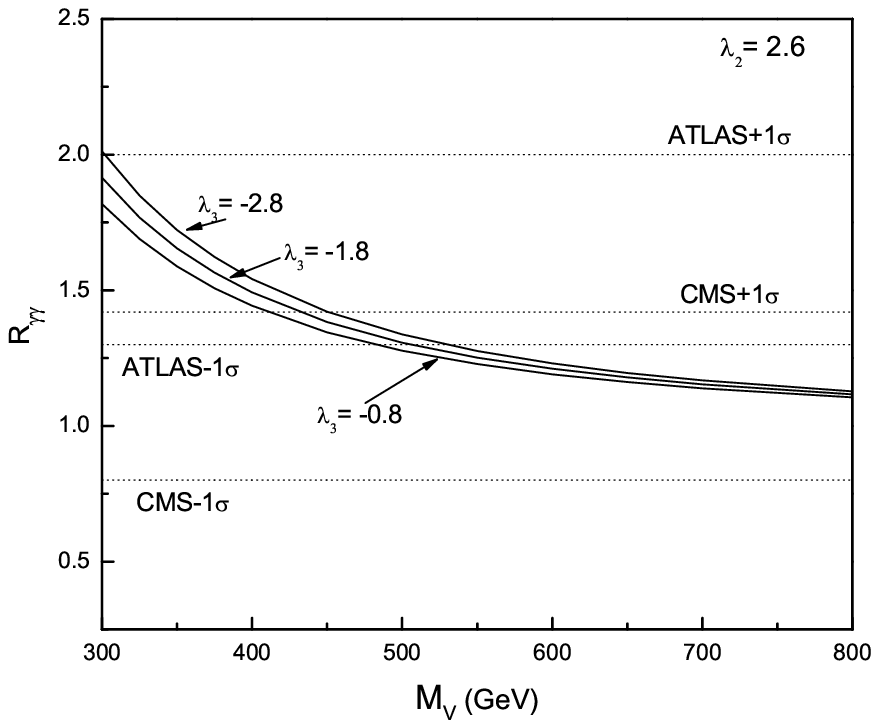}}
\caption{The parameter $R_{\gamma\gamma}$  as function of the mass $M_{V}$ for different values of the \hspace*{1.7cm} parameters $\lambda_{2}$ and $\lambda_{3}$.}\label{ee}
\vspace{-0.5cm}
\end{figure}

\begin{figure}[htb]
\vspace{-0.5cm}

{\includegraphics[scale=0.68]{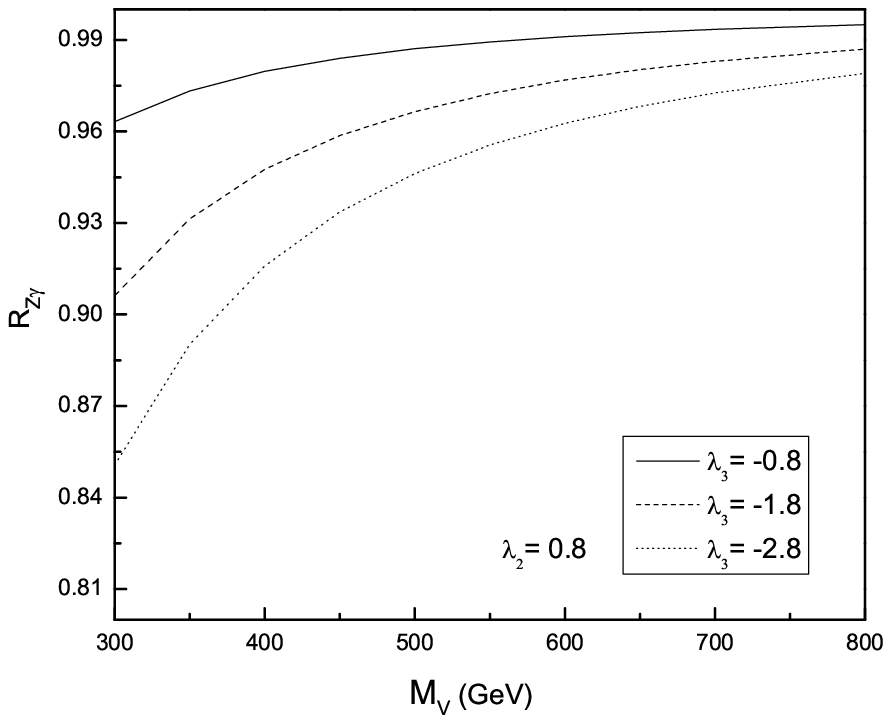}}
{\includegraphics[scale=0.68]{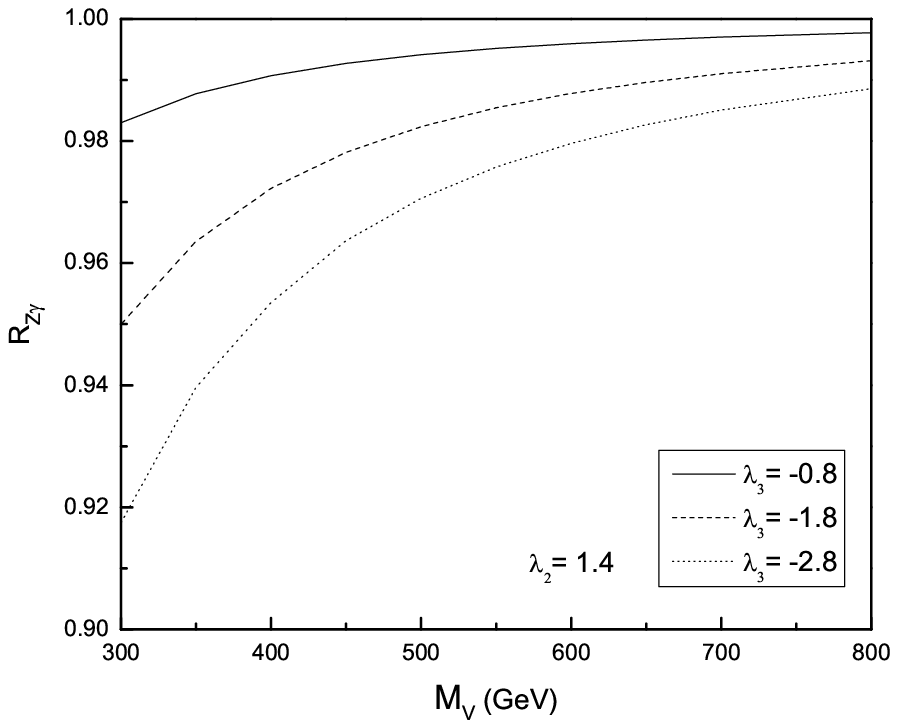}}

{\includegraphics[scale=0.68]{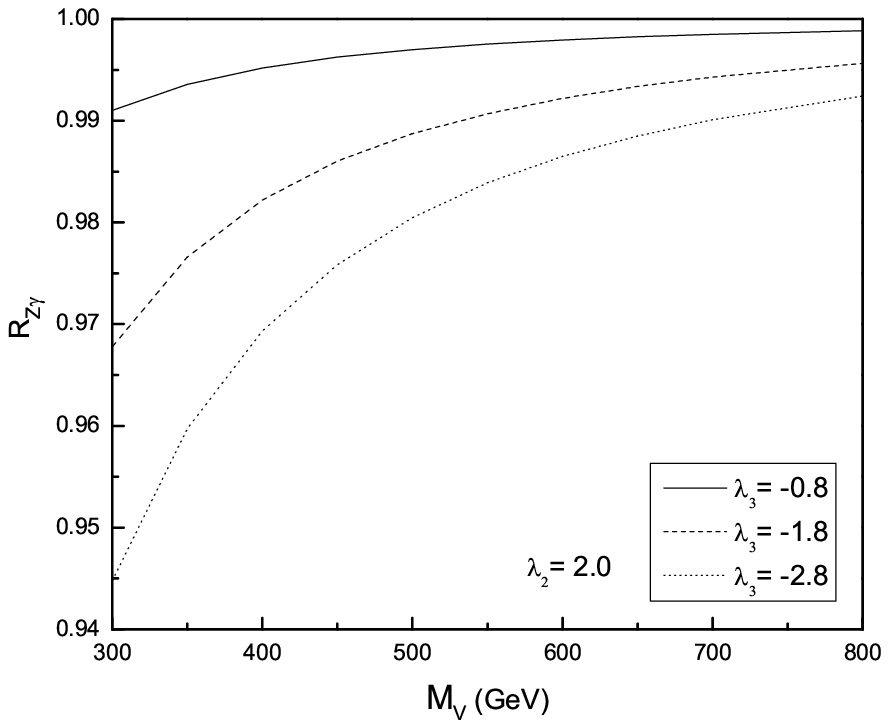}}
{\includegraphics[scale=0.68]{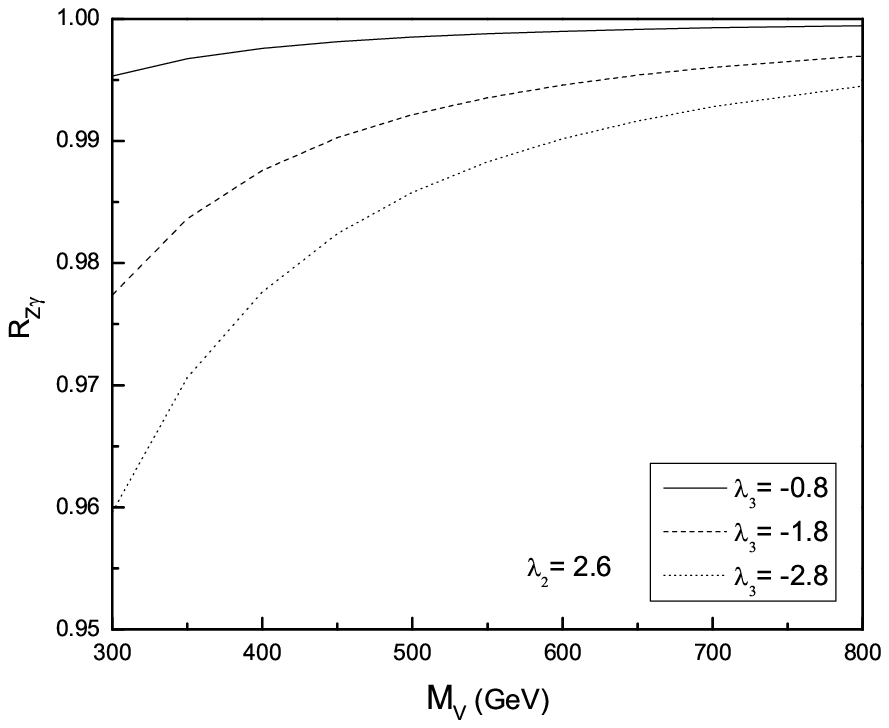}}
\vspace{-0.5cm}
\caption{The parameter $R_{Z\gamma}$ as function of the mass $M_{V}$ for different values of the  \hspace*{1.7cm} parameters $\lambda_{2}$ and $\lambda_{3}$.}\label{ee}
\end{figure}
In order to consider the contributions of the vector bileptons $V^{\pm}$ and $U^{\pm\pm}$ to the $LHC$ signal rates for the $\gamma\gamma$ and $Z\gamma$ channels, as usual, we define the so called $R$ parameters as
\begin{eqnarray}
R_{\gamma\gamma}=\frac{\sigma(p p \rightarrow h)Br(h \rightarrow \gamma \gamma)}{\sigma_{SM}(pp \rightarrow h)Br_{SM}(h \rightarrow \gamma\gamma)},
\end{eqnarray}
\begin{eqnarray}
R_{Z \gamma}=\frac{\sigma(p p \rightarrow h)Br(h \rightarrow Z \gamma)}{\sigma_{SM}(pp \rightarrow h)Br_{SM}(h \rightarrow Z\gamma)}.
\end{eqnarray}
In our case, the Higgs production rates are almost same as those for the $SM$.

Considering the expressions of the gauge boson masses, there are $M_{U}^{2}-M_{V}^{2}=m_{W}^{2}$ and $\nu_{\rho}/\nu_{\chi}\approx m_{W}/M_{V}$, so the parameters $R_{\gamma\gamma}$ and $R_{Z\gamma}$ mainly depend on the free parameters $M_{V}$, $\lambda_{2}$ and $\lambda_{3}$. In order to allow perturbative calculations, the dimensionless  parameters of the Higgs potential must be satisfy $-3 < \lambda_{i} < 3$. For the $RM$331 model there are $0 < \lambda_{2} < 3$ and $-3 < \lambda_{3} <0$ [13]. Our numerical results are summarized in Fig.3 and Fig.4, in which we plot the parameters $R_{\gamma\gamma}$ and $R_{Z\gamma}$ as functions of the mass $M_{V}$ for different values of the parameters $\lambda_{2}$ and $\lambda_{3}$. One can see from these figures that the vector bileptons generate positive contributions to the parameter $R_{\gamma\gamma}$, while give negative contributions to the parameter $R_{Z\gamma}$, resulting in the  anti-correlation between the widths $h \rightarrow \gamma \gamma$ and $h \rightarrow Z\gamma$, which could be tested in the future experiments at the $LHC$. All contributions decrease as the masses of vector bileptons increasing and there are $R_{\gamma\gamma}\rightarrow 1$ and $R_{Z\gamma}\rightarrow 1$ for $M_{V}\rightarrow \infty$.  Considering the constraint of the muon $g-2$ $a_{\mu}$ on the masses of the vector bileptons, i. e. $308GeV \leq M_{V} \leq 429GeV$, one can find that the vector bileptons predicted by the $RM331$ model might explain the $ALTAS$ data for the $\gamma\gamma$ signal in the $1\sigma$ range for $\lambda_{2} =0.8$, $-1.8< \lambda_{3} <0$, and for $\lambda_{2} =2$, $-2.8< \lambda_{3} <0$. While the $CMS$ data produce severe constraints on the parameters $ \lambda_{2}$ and $ \lambda_{3}$.

\noindent{\bf 6. Conclusions and discussions}

Vector bileptons are  one kind of new particles, which can generate special signals in high energy collider experiments. In this paper, we  consider the contributions of the vector bileptons $V^{\pm}$ and $U^{\pm\pm}$ predicted by the $RM331$ model to the muon $g-2$ $a_{\mu}$ and compare our numerical result with the discrepancy  between its experimental result and $SM$ prediction. It is found that the precise measured value of $a_{\mu}$ can give severe constraints on the masses of $V^{\pm}$ and $U^{\pm\pm}$. Taking into consideration of these constraints, we calculate  the contributions of the vector bileptons $V^{\pm}$ and $U^{\pm\pm}$ to the Higgs decay channels $h \rightarrow \gamma \gamma$ and $Z\gamma$ in the context of the $RM331$ model. Our numerical results show that  the vector bileptons enhance the partial width  $\Gamma(h\rightarrow \gamma\gamma)$, while reduce the partial width  $\Gamma(h\rightarrow Z\gamma)$, which are  anti-correlated. With reasonable values of the relevant free parameters, the vector bileptons can explain the $LHC$ data for the $\gamma\gamma$ signal. If the  $CMS$ data persists, the value of the parameters $ \lambda_{2}$ and $ \lambda_{3}$ should be severe constrained.

From section 2 we can see that, in addition to the vector bileptons $V^{\pm}$ and $U^{\pm\pm}$,  the $RM331$ model also predicts the existence of other new charged particles, such as new fermions and doubly charged scalars. These new fermions can not directly couple to the Higgs boson $h$, so they have no contributions to  the Higgs decay channels $h \rightarrow \gamma \gamma$ and $Z\gamma$ at leading order. Compare to the vector bileptons $V^{\pm}$ and $U^{\pm\pm}$, the doubly charged Higgs bosons $H^{\pm\pm}$ may contribute to these decay channels destructively [21], which might relax the constrains of the $LHC$ data on the $RM331$ model. However, $H^{\pm\pm}$ have masses around $TeV$ scale, their contributions to $h \rightarrow \gamma \gamma$ and $Z\gamma$ are much smaller than those of  vector bileptons.

\section*{Acknowledgments} \hspace{5mm}This work was
supported in part by the National Natural Science Foundation of
China under Grants No.11275088, the Natural Science Foundation of the Liaoning Scientific Committee
(No. 201102114), and Foundation of Liaoning Educational Committee (No. LT2011015).

\end{document}